\documentclass[preprint,12pt]{elsarticle}

\usepackage{amssymb}
\usepackage{amsmath}
\usepackage{amsfonts}
\usepackage{times}
\usepackage{url}
\usepackage{graphicx}
\usepackage{color}
\usepackage{subfigure}
\makeatletter
\renewcommand{\@thesubfigure}{}

\newcommand{\comment}[1]{}
\definecolor{Orange}{rgb}{1,0.5,0}

\makeatletter
\renewcommand{\@thesubfigure}{}

\newcommand{\Equation}[1]{Eqn.~(#1)}

\begin{document}

\begin{frontmatter}

\title{Image Tag Refinement by Regularized Latent Dirichlet Allocation}

\author[msra]{Jingdong Wang \corref{cor1}}
\ead{jingdw@microsoft.com}
\ead[url]{http://research.microsoft.com/en-us/um/people/jingdw/}
\author[jiazhen]{Jiazhen Zou}
\ead{jiazhenzhou@yahoo.com}
\author[xuhao]{Hao Xu}
\ead{xuhao657@gmail.com}
\author[msra]{Tao Mei}
\ead{tmei@microsoft.com}
\author[msrr]{Xian-Sheng Hua}
\ead{xshua@microsoft.com}
\author[msra]{Shipeng Li}
\ead{spli@microsoft.com}

\cortext[cor1]{Corresponding author}

\address[msra]{Microsoft Research Asia, Beijing, P.R. China}
\address[jiazhen]{Columbia University, New York, USA}
\address[xuhao]{University of Science and Technology of China, Hefei, P.R. China.}
\address[msrr]{Microsoft Research Redmond, Redmond, USA}

\begin{abstract}
Tagging is nowadays the most prevalent and practical way to make images searchable.
However,
in reality many manually-assigned tags are irrelevant to image content
and hence are not reliable for applications.
A lot of recent efforts have been conducted to refine image tags.
In this paper,
we propose to do tag refinement from the angle of topic modeling
and present a novel graphical model,
regularized Latent Dirichlet Allocation (rLDA).
In the proposed approach,
tag similarity and tag relevance are jointly estimated in an iterative manner,
so that they can benefit from each other,
and the multi-wise relationships among tags
are explored.
Moreover,
both the statistics of tags and visual affinities of images in the corpus
are explored
to help topic modeling.
We also analyze the superiority of our approach
from the deep structure perspective.
The experiments on tag ranking and image retrieval
demonstrate the advantages of the proposed method.
\end{abstract}


\end{frontmatter}

%
%


\section{Introduction}
The community-contributed multimedia content
in the internet,
such as Flickr, Picasa, Youtube and so on,
has been exploding.
To facilitate the organization of the uploaded images or videos,
media repositories usually offer a tool
to enable consumers
to manually assign tags (a.k.a. labels)
to describe the media content~\cite{AmesN07}.
These assigned tags are adopted
to index the images
to help consumers access shared media content.

\begin{figure}[t]
\begin{centering}
\includegraphics[width=0.4\textwidth]{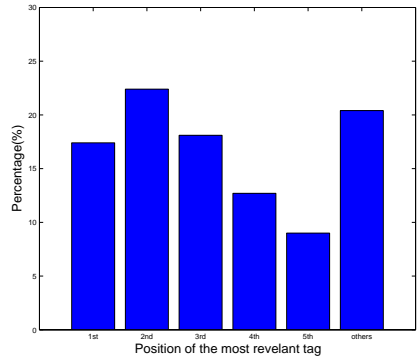}
\caption{A statistic on the MSRA-TAG dataset~\cite{LiuHYWZ09}
indicating the percentage of images
with the most relevant tag
in different positions.}
\label{fig:TagExample}
\end{centering}
\end{figure}

Reliable tagging results in
making shared media more easily accessible
to the public.
However, the reliability of tagging
is not guaranteed
in that the tags may be noisy,
orderless and incomplete~\cite{KennedyCK06},
possibly due to carelessness of the taggers.
First,
some tags are noises and may be irrelevant to media.
According to the statistics in Flickr,
there are about only $50\%$ tags
indeed relevant to photos~\cite{KennedyCK06, ChuaTHLLZ09}.
Second, different tags essentially have different relevance degrees
to the media,
but such information is not indicated in the current tag list,
where the order is given
according to the input sequence.
We did an analysis on the MSRA-TAG dataset~\cite{LiuHYWZ09},
which was crawled from Flickr, about what percentage of images
have the most important tags
in different positions.
A statistics figure
is shown in~Fig.{\ref{fig:TagExample}}
to indicate the result.
It can be observed from this statistics
that less than $20\%$ images have the most relevant tags
at the top position,
which shows that the tags are almost in a random order
in terms of the relevance.
Last, the tags of some photos are incomplete
due to the interest limitation of taggers,
and even not given.

We address the problem of
refining the tags,
to facilitate the access of the shared media.
To be specific, we investigate the tagging problem
in Flickr, one of the most popular photo sharing web sites,
and propose to reorder the tags.
The available information to refine the tags
consists of manual tags and image affinity.
\begin{enumerate}
  \item Although they are not completely reliable,
  the manual tags still reflect the photo content
  in some degree
  and their relations can be explored
  for tag refinement.
  Existing solutions only make use of
  the pairwise relation between tags,
  mined from WordNet~\cite{SigurbjornssonZ08},
  or estimated from Web photo tags~\cite{LiSW08,LiSW09,LiuHYWZ09,WeinbergerSZ08}.

  \item Visually similar images usually tend to
  have similar semantics and hence have similar tags,
  which means that the tag refinement of one image may
  benefit from those of other images.
  The typical exploration~\cite{LiuHYWZ09,LiSW09} is to utilize the visual popularity of one image
  among the images having the same tag
  as a cue to estimate the relevance of the tag with this image.
\end{enumerate}

\begin{figure*}
\centering
\subfigure[(a)]{\label{fig:IllustrationByDeeplearning:pairwise}\includegraphics[scale=1]{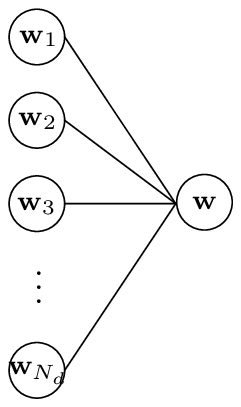}}~~~
\subfigure[(b)]{\label{fig:IllustrationByDeeplearning:lda}\includegraphics[scale=1]{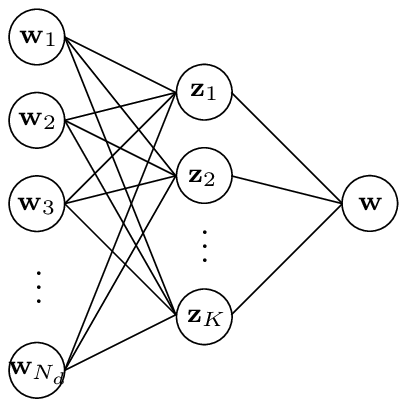}}~~~
\subfigure[(c)]{\label{fig:IllustrationByDeeplearning:rlda}\includegraphics[scale=1]{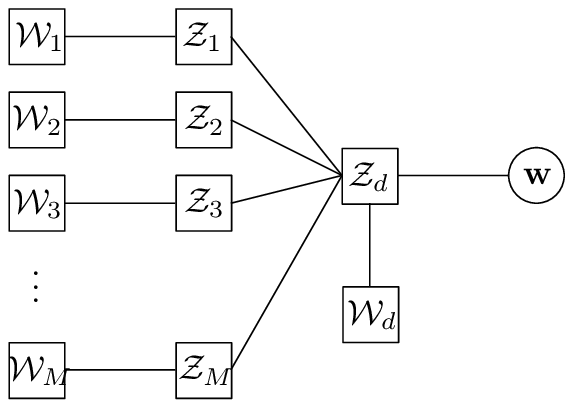}}
\caption{Illustration from the graphical representations.
(a) Two layers for pairwise based approaches.
(b) Three layers for LDA.
(c) Four layers for our approach (rLDA).
It can be concluded that
our approach is deeper.}
\label{fig:IllustrationByDeeplearning}
\end{figure*}

In this paper,
we present a novel probabilistic formulation,
to estimate the relevance of a tag
by considering all the other images and their tags.
To this goal, we propose a novel model
called regularized latent Dirichlet allocation (rLDA),
which estimates the latent topics
for each document,
with making use of other documents.
The model is applicable in tag refinement
due to the observation that
the content of an image essentially
contains a few topics
and the reasonable assumption that the tags assigned to the image
accordingly form a few groups.
The latent topics are estimated
by viewing the tags of each image as a document,
and the estimation also benefits
from other visually similar images
by the regularization term,
instead of the estimation by LDA
only from the corresponding document.
The main contribution of our approach lies
in the following aspects.
On the one hand, both LDA and rLDA explore the multiple-wise relation
among tags through the latent topics,
rather than pairwise relations in the random walk based methods.
On the other hand,
the tag relevance estimation from rLDA can be interpreted
using the deep structure~\cite{Bengio09, HintonOT06}.
Compared with random walk and LDA,
our approach is the deepest,
and the illustration is presented
in Figure {\ref{fig:IllustrationByDeeplearning}}.

\section{Related work}
The automatic image tagging or annotation problem is usually regarded
as an image classification task.
Typical techniques~\cite{BarnardDFFBJ03, BleiJ03, CarneiroCMV07, ChenCCTHW08, DattaJLW08, FengL08, JeonM04, JinCS04, LiW03, NguyenKPT10, PutthividhyaAN10, WangHLZYC12}
usually learn a generative/discriminative multi-class classifier
from the training data,
to construct a mapping function
from low level features extracted from the images
to tags,
and then predict the annotations or tags for the test images.
Later, a more precise formulation
is presented to regard it
as a multi-label classification problem
by exploring the relations
between multiple labels~\cite{JiangCL06, QiHRTMZ07}.
The automatic annotation techniques have shown
great successes
with small scale tags
and the well-labeled training data.
But in the social tags,
e.g., image tags on Flickr,
there exist noisy or low-relevance tags,
and the vocabulary of tags is very large,
which limits the performance of conventional automatic tagging techniques
in social tagging.
The study in~\cite{KennedyCK06}
has shown that
classifiers trained
with Flickr images
and associated tags
got unsatisfactory performance
and that
tags provided by Flickr users
actually contain noise.
Moreover,
the relevance degrees of the tags,
i.e., the order of the tags,
are not investigated
in automatic annotation.

Various approaches have been developed
to refine tags
using the available tags and visual information.
The following reviews some closely-related methods,
and more discussions can be found from a survey~\cite{WangNHC12}.
The straightforward approach
directly exploits the tag relation,
e.g., co-occurrence relation mined from WordNet~\cite{Miller95}, or the internet,
and then refines tags~\cite{SigurbjornssonZ08, WeinbergerSZ08, KrestelFN09}.
For example, the tag ambiguities are resolved~\cite{WeinbergerSZ08}
by finding two tags that appear in different contexts but
are both likely to co-occur with the original tag set
and then presenting such ambiguous tags to users
for further clarification.
The random walk approach
over the pairwise graph
on the provided tags
with edges weighted by the tag similarities
is presented in~\cite{WangJZZ06, LiuHYWZ09}.
The visual information is proved very useful
to help tag refinement.
For example, the neighborhood voting approach~\cite{LiSW08} is
to recommend the tags
by exploring the tags of the visually similar images.
The likelihood that a tag is associated with an image
is computed in~\cite{WangJZZ06, LiuHYWZ09}
from probabilistic models learnt the images assigned with such a tag,
and then put it into the random walk framework
for further refinement.
A hybrid probabilistic model~\cite{ZhouCQX11}
is introduced
to combine both collaborative and content based algorithms
for tagging,
which is similar to~\cite{LiuHYWZ09}
in using the visual contents.
A RankBoost based approach~\cite{WuYYH09} is presented
to learn a function to combine ranking features from multi-modalities,
including tag and visual information.
An optimization framework~\cite{LiuHWZ10}
is proposed
to perform tag filter and enrichment
by exploring visual similarity
and additional knowledge from WordNet~\cite{Miller95}.
An approach~\cite{ZhuYM10} formulates the tag refinement problem
as a decomposition of the user-provided tag matrix
into a low-rank matrix and a sparse error matrix,
targeting the optimality
by low-rank, content consistency,
tag correlation and error sparsity.

Rather than exploring the pairwise relation among tags,
some techniques are proposed to
adopt the multiple wise relations among tags,
through latent models.
Latent topic models,
alternatives of latent Dirichlet allocation,
is adopted~\cite{KrestelFN09, KrestelF09, BundschusYTRD09} to
learn a generative model from the tags,
which then can estimate the posterior probability
that a tag is associated with an image.
Those methods are limited in
lack of capabilities of adopting visual information.
Therefore, this paper proposes a novel topic model,
called regularized latent Dirichlet allocation,
to estimate the topic models
with exploiting the visual information.
The latent topic based models are also justified
by the conclusion in~\cite{SigurbjornssonZ08}
that
these tags assigned to images span a broad spectrum of the semantic space.

From the perspective of the deep learning theory~\cite{HintonOT06},
the random walk based approaches essentially
estimate the tag relevance
with a shallow structure,
which only consists of two levels,
the provided tags as the input level
and the tag being considered as the output level.
The LDA based approach is with a deep structure,
introducing a latent topic level,
which has potential to get better performance.
The proposed regularized LDA model is deeper,
with four levels,
the tags associated with other images
as the first level,
the latent topics of other images
and the tags of the image being considered
as the second level,
the latent topic as the third level,
and the tag being considered as the output level.

The relational topic model~\cite{ChangB09}
is closely related to the proposed regularized LDA.
But they are clearly different
because our approach imposes the regularization
over the topic distribution instead of the latent variables
and moreover our approach deals multiple modalities
and makes use of additional visual similarity
to formulate the regularization term.
Our approach is also different from topic models for image annotation~\cite{BleiJ03, BarnardDFFBJ03, NguyenKPT10, PutthividhyaAN10}:
The image tagging problem in our paper is more challenging than image annotation as aforementioned.
and moreover the proposed regularized LDA aims to impose the consistency
of tags within similar images while they are supervised algorithms~\cite{NguyenKPT10, PutthividhyaAN10}
or aim to find common topics shared by tags and visual contents.
This paper is different from the short version~\cite{XuWHL09}
because we present a formal derivation of our approach,
introduce a new inference approach
conduct more experiments,
and particularly,
we use the deep network structure to analyze
the benefit of regularized LDA.

\section{Taxonomy}
The input consists of an image corpus,
$M$ images $\mathcal{I} = \{I_1, \cdots, I_M\}$,
and $M$ tag documents $\mathcal{W} = \{\mathcal{W}_1, \cdots, \mathcal{W}_M\}$,
with $\mathcal{W}_k$ being the set of tags manually assigned
to image $I_k$.
Here $\mathcal{W}_k$ is defined, similarly to~\cite{BleiNJ03},
as a sequence of $N_k$ words
denoted by $\mathcal{W}_k = \{\mathbf{w}_1, \mathbf{w}_2, \cdots, \mathbf{w}_{N_k}\}$.
$\mathbf{w}_i$ represents a word,
an item from a vocabulary indexed by $\{1, \cdots, V\}$,
and is a $V$-dimensional vector, with only one entry equal to 1
and all the other entries equal to 0,
e.g., $w^v_i = 1$ and $w^u_i = 0$ for $u \neq v$
if $\mathbf{w}_i$ represents the $v$-th word.
Besides, we use $\mathbf{f}_k$
to represent the visual feature of image $I_k$.
The goal is to reorder the tags in each document,
finding reordered tags $\{\mathbf{w}_{n_1}, \mathbf{w}_{n_2}, \cdots, \mathbf{w}_{n_{N_k}}\}$
with $\{n_1, n_2, \cdots, n_{N_k}\}$ being
a permutation of $\{1, 2, \cdots, N_k\}$,
so that the tags on the top are semantically
more relevant to the image.

Given the image corpus and the associated tag documents,
we introduce a term, tag relevance,
which is then used to reorder the tags.
The tag relevance is inferred
based on the joint probability,
$P(\mathcal{W}_1, \cdots, \mathcal{W}_d, \cdots, \mathcal{W}_M | I_1, \cdots, I_d, \cdots, I_M)$.

\subsection{Formulation with the tag cue}
The manually-assigned tags,
in some degree,
describe the semantic content of an image,
although it may contain noise,
or not be complete.
Hence,
as a candidate solution,
the relevance of each tag
can be inferred
from these tags.
The joint probability over the tags $\mathcal{W}_d$
of image $I_d$ is formulated as
a pairwise Markov random field (MRF),
\begin{align}
P(\mathcal{W}_d)
\propto \prod_{i, j \in \{1, \cdots, N_d\}} P(\mathbf{w}_{di}, \mathbf{w}_{dj}),
\end{align}
where $P(\mathbf{w}_{di}, \mathbf{w}_{dj})$ is valuated from the tag relation.
This model can be further interpreted
by a random walk model.
Specifically, a transition probability
is defined from the tag relation as
\begin{align}
T_{\mathbf{w}_{di} \rightarrow \mathbf{w}_{dj}} \propto \text{sim}(\mathbf{w}_{di}, \mathbf{w}_{dj}).
\label{eqn:transitionprobability}
\end{align}
Here $T_{\mathbf{w}_{di} \rightarrow \mathbf{w}_{dj}}$ has been normalized
such that $\sum_{j \in \{1, \cdots, N_d\}}T_{\mathbf{w}_{di} \rightarrow \mathbf{w}_{dj}} = 1$.
$\text{sim}(\mathbf{w}_{di}, \mathbf{w}_{dj})$ is the similarity
between $\mathbf{w}_{di}$ and $\mathbf{w}_{dj}$,
and may be computed from WordNet~\cite{SigurbjornssonZ08} or other methods~\cite{LiSW08,LiSW09,LiuHYWZ09,WeinbergerSZ08}.
Given this model,
the stable distribution of this model,
$\mathbf{p}^s =[ P(\mathbf{w}_{d1})~\cdots~P(\mathbf{w}_{dN_d})]$
is then used to evaluate the relevance of each tag.

\subsection{Formulation with the visual cue}
The visual description of a tag,
$u$, can be obtained
from a set of images,
$\mathcal{I}_u = \{I_k | u \in \mathcal{W}_k \}$,
associated with that tag.
Given an image $I$,
the posteriori probability
$p(u|I)$
can be computed as follows,
\begin{align}
p(u|I) = \frac{p(I|u)p(u)}{p(I)} \propto p(I|u)p(u),
\end{align}
where $p(I|u)$ can be estimated by
$p(I|u) = p(I|\mathcal{I}_u)$
that can be computed using the kernel density estimation~\cite{LiuHYWZ09},
e.g., $p(I|u) \propto \sum_{I' \in \mathcal{I}_u} K(I, I')$.
The scheme estimating the density
can also be interpreted as
the stable distribution
of a random walk over the images $\mathcal{I}_u$,
where the transition probability is estimated
from the kernel $K(I, I')$.
Without any bias for any tag,
$p(u)$ can be thought as a uniform distribution.
To the end,
$p(u|I) \propto p(I|u)$ can be used as the relevance
of tag $u$ for image $I$.

\subsection{Formulation with both tag and visual cues}
As a straightforward scheme exploring
of both tag and visual cues,
the relevances from the visual cue
can be viewed as observations
to the probability model over tags.
Denote $p_v(\mathbf{w}_{di})$ as the observations of tag $\mathbf{w}_{di}$,
the model can be written as
\begin{align}
P(\mathcal{W}_d)
\propto \prod_i p_v(\mathbf{w}_{di}) \prod_{i, j \in \{1, \cdots, N_d\}} P(\mathbf{w}_{di}, \mathbf{w}_{dj}).
\end{align}
It can also be interpreted as a random walk with restarts.
The transition model is the same to~\Equation{\ref{eqn:transitionprobability}},
and $p_v(\mathbf{w}_{di})$, first normalized so that
$\sum_{i \in \{1, \cdots, N_d\}} p_v(\mathbf{w}_{di}) = 1$,
is viewed as the restarts.
The resulted approach includes two steps:
relevance from visual cue
and a random walk with restart over tags.
The solution is essentially the fixed point
$\mathbf{p}^s = \lambda \mathbf{T}^T\mathbf{p}^s + (1-\lambda)\mathbf{p}^v$,
where $\mathbf{p}^v = [p_v(\mathbf{w}_{d1})\cdots p_v(\mathbf{w}_{dN_d})]$.
The two-step approach,
presented in~\cite{LiuHYWZ09},
can be cast into this formulation.

\subsection{Extension to a joint formulation}
The above formulations actually consider each image independently,
and as a result,
tag relevances for each image can be estimated separately.
In the computation of $p(I|u) = p(I|\mathcal{I}_u)$,
each image in the set,
$\mathcal{I}_u = \{I_k | u \in \mathcal{W}_k \}$,
is equivalently considered.
In fact,
an image is associated with several tags,
and accordingly the typicality degree
of each image to represent the tag
may be different.
Alternatively,
the tag relevance can be jointly estimated
by considering a joint probability,
\begin{align}
& P(\mathcal{W}_1, \cdots, \mathcal{W}_d, \cdots, \mathcal{W}_M) \nonumber \\
\propto & (\prod_d \prod_{u, v \in \{1, \cdots, N_d\}} P(\mathbf{w}_{du}, \mathbf{w}_{dv}))
( \prod_d\prod_u p(I_d | u)),
\end{align}
where the first term corresponds to the pairwise MRF,
and the second term is from the visual constraint
and can be viewed as the visual regularization.
$p(I|u)$,
can be furthermore written as
$p(I|u) = p(I|\mathcal{I}_u, \mathcal{A}_u)
\propto \sum_{I' \in \mathcal{I}_u} a_u(I')K(I, I')$,
(viewed as a weighted random walk model)
where $\mathcal{A}_u = \{a_u(I')|I' \in \mathcal{I}_u\}$ is a set of weights,
with each corresponding to the typicality degree of each image
and estimated as $P(I|u)$.
To the end, the solution of the approach can be equivalently obtained
by solving a fixed point problem,
\begin{align}
\mathbf{p}^s_{d\cdot} &= \lambda \mathbf{T}_d^T \mathbf{p}^s_{d\cdot} + (1 - \lambda) \mathbf{p}^o_{d\cdot}, \\
\bar{\mathbf{p}}^o_{\cdot u} &= \mathbf{T}^w_u \bar{\mathbf{p}}^o_{\cdot u},
\end{align}
where $\mathbf{T}^w_u \propto \operatorname{diag}(\mathbf{p}^s_{\cdot u}) \mathbf{P}$ is a weighted transition matrix,
with $\mathbf{P}$ corresponding to the pairwise probability $P(\mathbf{w}_{du}, \mathbf{w}_{dv})$,
$\bar{\mathbf{p}}^o_{\cdot u}$ and $\mathbf{p}^o_{d\cdot}$
are the observation probability vectors,
a column vector and a row vector
of $\mathbf{P}=[p(I_d |u)]_{M\times V} $
normalized
w.r.t. image set and tag set, respectively.

\section{Our approach}
The aforementioned approach builds the relationship
among the tags,
using the pairwise framework,
which is lack of the ability
to describe the multiple wise relations
among tags.
However, we observed that the relation among tags
is beyond pairwise and
that some tags may have closed relations
and the tags associated with one image form several meaningful cliques.
To make use of the multiple wise relations,
we introduce a topic model
to construct the tag relation,
which uses latent topic variables
to connect the tags
for building the multiple wise relationships.
Moreover, rather than separately building the topic model
independently for each image,
we propose a novel approach
to jointly model the images together
based on the observation that
visually similar images should have similar semantic contents,
To this goal,
we introduce smoothness terms
over the latent topics of images.

\subsection{Tag refinement via latent Dirichlet allocation}
\begin{figure}
\centering
\subfigure[(a)]{\label{fig:lda:lda}\includegraphics[scale = 1.2]{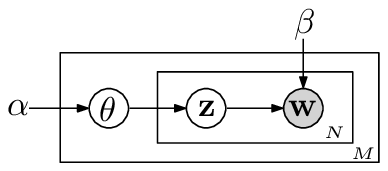}}~~~~
\subfigure[(b)]{\label{fig:lda:variational}\includegraphics[scale = 1.2]{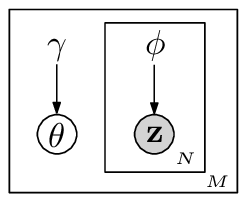}}
\caption{(a) Graphical model representation of LDA.
(b) Graphical model representation
of the variational distribution used to approximate the posterior in LDA}
\label{fig:lda}
\end{figure}

\subsubsection{Latent Dirichlet allocation}
Latent Dirichlet allocation (LDA) is a generative probabilistic model
of a corpus, e.g., a set of documents.
The basic idea is that
documents are represented as random mixtures
over latent topics,
where each topic is characterized by a distribution over tags (words),
or intuitively is viewed as
a group of soft (partially weighted) tags.
LDA is a special directed graphical model (a.k.a. Bayesian network),
and its graphical representation is shown in~Figure~\ref{fig:lda:lda}.
In this graphical model,
$\mathbf{z}$ is a topic vector
of length $k$,
with $k$ being the number of latent topics,
where it corresponds to the $i$ topic
if $z^i = 1$ and $z^j = 0$ for $j \neq i$.
$\boldsymbol{\theta}$ is a $k$-dimensional vector,
the parameter of the multinomial distribution.
$\boldsymbol{\alpha}$ is the Dirichlet parameter,
a vector of $k$ positive reals.
Matrix $\boldsymbol{\beta}$, of dimension $k \times V$,
is used to parameterize the tag-topic probability,
where $\beta_{ij} = p(w^j = 1 | z^i = 1)$.

It can be interpreted as a generative process for each document $\mathcal{W}_d$
as follows.
\begin{enumerate}
  \item Choose $\boldsymbol{\theta}_d \sim \operatorname{Dir}(\boldsymbol{\alpha})$.
  \item For each of the $N_d$ tags $\mathbf{w}_{dn}$,
  \begin{enumerate}
    \item Choose a topic $\mathbf{z}_{dn} \sim \operatorname{Multinomial}(\boldsymbol{\theta}_d)$.
    \item Choose a tag $\mathbf{w}_{dn}$ from $p(\mathbf{w}_{dn} | \mathbf{z}_{dn}; \boldsymbol{\beta})$,
    a multinomial probability conditioned on the topic $\mathbf{z}_{dn}$.
  \end{enumerate}
\end{enumerate}

In the graphical representation,
there are three levels.
The parameters $\boldsymbol{\alpha}$ and $\boldsymbol{\beta}$
are corpus level parameters,
assumed to be sampled once
in the generative process
for a single corpus.
The variable $\boldsymbol{\theta}$
is a document-level (image-level) variable
sampled for each document (image).
Finally, the variables $\mathbf{z}_{dn}$
and $\mathbf{w}_{dn}$
are word-level (tag-level) variables,
sampled once for each word in each document.

\subsubsection{Topic distribution}
The goal to use LDA here is to mine
the latent topic distribution for a given document $\mathcal{W}_d$,
$p(\boldsymbol{\theta} | \mathcal{W}_d, \boldsymbol{\alpha}, \boldsymbol{\beta})$.
The topic distribution can be obtained
by integrating out other latent variables $\mathbf{z}_1, \cdots, \mathbf{z}_{N_d}$
over the posterior distribution,
$p(\boldsymbol{\theta}, \{\mathbf{z}_1, \cdots, \mathbf{z}_{N_d}\} | \mathcal{W}_d, \boldsymbol{\alpha}, \boldsymbol{\beta})$.
This distribution is intractable to compute in general.
The approximate inference algorithms,
e.g., Markov chain Monte Carlo and variational inference,
can be used to tackle this problem.
The variational inference algorithm introduces
a variational distribution,
\begin{align}
q(\boldsymbol{\theta}, \{\mathbf{z}_1, \cdots, \mathbf{z}_{N_d}\} | \boldsymbol{\gamma}, \{\boldsymbol{\phi}_1, \cdots, \boldsymbol{\phi}_{N_d}\})
= q(\boldsymbol{\theta} | \boldsymbol{\gamma}) \prod_{n=1}^{N_d} q(\mathbf{z}_n | \boldsymbol{\phi}_n), \nonumber
\end{align}
where the Dirichlet parameter $\boldsymbol{\gamma}$
and the multinomial parameters $\{\boldsymbol{\phi}_1, \cdots, \boldsymbol{\phi}_{N_d}\}$
are the free variational parameters.
The variational parameters
can be obtained
from the following optimization problem:
\begin{align}
& \boldsymbol{\gamma}*, \{\boldsymbol{\phi}*_1, \cdots, \boldsymbol{\phi}*_{N_d}\} \nonumber \\
= & \arg \min_{\boldsymbol{\gamma}, \{\boldsymbol{\phi}_1, \cdots, \boldsymbol{\phi}_{N_d}\}} \nonumber\\
& \operatorname{KL}(q(\boldsymbol{\theta}, \{\mathbf{z}_1, \cdots, \mathbf{z}_{N_d}\} | \boldsymbol{\gamma}, \{\boldsymbol{\phi}_1, \cdots, \boldsymbol{\phi}_{N_d}\})
\nonumber\\
&~~~~~ ||  p(\boldsymbol{\theta}, \{\mathbf{z}_1, \cdots, \mathbf{z}_{N_d}\}| \mathcal{W}_d, \boldsymbol{\alpha}, \boldsymbol{\beta})),
\end{align}
where $\operatorname{KL}(q(\cdot) | p(\cdot))$
is the Kullback-Leibler (KL) divergence
between the variational distribution $q(\cdot)$
and the true posterior $p(\cdot)$.
Figure~\ref{fig:lda:variational} illustrates this variational distribution.
This minimization can be achieved
via an iterative fixed-point method.
The update equations are as follows,
\begin{align}
\phi_{ni} & \propto \beta_{iv(\mathbf{w}_n)} \exp\{\operatorname{E}_q[\log (\theta_i) |\boldsymbol{\gamma}]\} \\
\gamma_i & = \alpha_i + \sum_{n=1} ^N \phi_{ni},
\end{align}
where $v(\mathbf{w})$ is a function that maps a vector representation of a word
to an index in the vocabulary.

In the LDA model,
two model parameters,
$\boldsymbol{\alpha}$ and $\boldsymbol{\beta}$,
can be estimated
from the given corpus of documents,
$\mathcal{W}$,
by maximizing the following log likelihood,
\begin{align}
L(\boldsymbol{\alpha}, \boldsymbol{\beta})
= \sum_{d = 1}^M \log p(\mathcal{W}_d | \boldsymbol{\alpha}, \boldsymbol{\beta}).
\end{align}
Here, $p(\mathcal{W}_d | \boldsymbol{\alpha}, \boldsymbol{\beta})$
can be efficiently estimated
by an expectation-maximization (EM) algorithm~\cite{BleiNJ03}.

\subsubsection{Tag relevance}
Given this topic model,
the relevance of a tag $\mathbf{w}$ for each image
is formulated as the probability
conditioned on the set of tags $\mathcal{W}$
associated with this image.
It is mathematically formulated as
\begin{align}
p(\mathbf{w} | \mathcal{W}) & =
\sum_{\bar{\boldsymbol{z}}} p(\mathbf{w}, \bar{\boldsymbol{z}} | \mathcal{W}) \nonumber\\
& = \sum_{\bar{\boldsymbol{z}}} p(\mathbf{w} | \bar{\boldsymbol{z}}) p(\bar{\boldsymbol{z}} | \mathcal{W}) \nonumber\\
& = \sum_{\bar{\boldsymbol{z}}} p(\mathbf{w} | \bar{\boldsymbol{z}}) \int p(\bar{\boldsymbol{z}}, \boldsymbol{\theta} | \mathcal{W}) d \boldsymbol{\theta} \nonumber\\
& = \sum_{\bar{\boldsymbol{z}}} p(\mathbf{w} | \bar{\boldsymbol{z}}) \int p(\bar{\boldsymbol{z}} | \boldsymbol{\theta}) p(\boldsymbol{\theta} | \mathcal{W}) d \boldsymbol{\theta} \nonumber\\
& \approx \sum_{\bar{\boldsymbol{z}}} p(\mathbf{w} | \bar{\boldsymbol{z}}) \int p(\bar{\boldsymbol{z}} | \boldsymbol{\theta}) q(\boldsymbol{\theta} | \boldsymbol{\gamma}) d \boldsymbol{\theta}.
\end{align}
The computation of this conditional distribution
can be illustrated by a graphical model
shown in~Figure~\ref{fig:IllustrationByDeeplearning:lda}.
From the analysis,
it can be observed that
the relations between tags are built
using the latent variables
that is beyond the pairwise relation,
illustrated in~Figure~\ref{fig:IllustrationByDeeplearning:pairwise},
and can capture the group information.

\subsection{Regularized latent Dirichlet allocation}
In the LDA model discussed above,
the distribution of topics
for each image is estimated separately.
However,
the tags associated with one image
may be incomplete and noisy.
Consequently,
the distribution of topics is not well estimated,
which influences the relevance of tags.
It is observed that
visually similar images usually have the same semantic content.
To utilize this property,
we introduce a regularization term
over the semantic content.
Rather than imposing this over tags directly,
we impose it over the latent topics
because
tags are sometimes too ambiguous for specifying a concept,
while topics usually have clearly conceptual meanings.

\begin{figure}
\centering
\includegraphics[scale = .8]{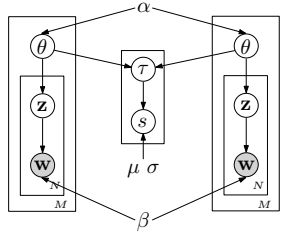}
\caption{The graphical representation of regularized LDA.}
\label{fig:rlda}
\end{figure}

The straightforward solution
to impose the regularization over topics
is a two-step sequential scheme:
first estimate the distribution of topics for each image,
and then to smooth the distribution
by considering the distributions of visually similar images.
Instead,
we propose a collective inference scheme
to estimate the distribution of latent topics.
To this goal, we build a joint distribution
over all the images,
called regularized latent Dirichlet allocation (rLDA).
This joint distribution is shown in Figure~\ref{fig:rlda}.
Different from the latent Dirichlet allocation model,
the topics over different images
are connected by an extra regularization model,
which is defined over
the topics of a pair of images.

It can be interpreted as a generative process
over the documents as follows.
\begin{enumerate}
  \item For each of the $M$ documents $\mathcal{W}_d$
      \begin{enumerate}
        \item Choose $\boldsymbol{\theta}_d \sim \operatorname{Dir}(\boldsymbol{\alpha})$.
        \item For each of the $N_d$ tags $\mathbf{w}_{dn}$,
       \begin{enumerate}
        \item Choose a topic $\mathbf{z}_{dn} \sim \operatorname{Multinomial}(\boldsymbol{\theta}_d)$.
        \item Choose a tag $\mathbf{w}_{dn}$ from $p(\mathbf{w}_{dn} | \mathbf{z}_{dn}; \boldsymbol{\beta})$,
        a multinomial probability conditioned on the topic $\mathbf{z}_{dn}$.
       \end{enumerate}
      \end{enumerate}

  \item For each of the set of document pairs $(\mathcal{W}_d, \mathcal{W}_{d'})$
      \begin{enumerate}
        \item Choose $\boldsymbol{\tau}_{dd'} \sim \operatorname{Multinomial}(\operatorname{R}(\boldsymbol{\theta}_d, \boldsymbol{\theta}_{d'}))$
        \item Choose a visual similarity $s_{dd'} \sim p(\mathbf{s}_{dd'} | \boldsymbol{\tau}_{dd'}; \boldsymbol{\mu}, \boldsymbol{\sigma})$,
        a Gaussian probability conditioned on the latent relation topic $\boldsymbol{\tau}_{dd'}$.
      \end{enumerate}
\end{enumerate}

Given the parameters,
$\boldsymbol{\alpha}$, $\boldsymbol{\beta}$,
$\boldsymbol{\mu}$ and $\boldsymbol{\sigma}$,
the joint distribution is given by
\begin{align}
&p(\{\boldsymbol{\theta}_d\}, \{\{\mathbf{w}_{dn}, \mathbf{z}_{dn}\}_n\}_d,
\{\mathbf{s}_{dd'}, \boldsymbol{\tau}_{dd'}\}_{dd'} |
\boldsymbol{\alpha}, \boldsymbol{\beta}, \boldsymbol{\mu}, \boldsymbol{\sigma}) \nonumber\\
= & \prod_{d=1}^{M} [ p(\boldsymbol{\theta}_d | \boldsymbol{\alpha})
\prod_{i=1}^{N_d}p(\mathbf{z}_{dn} | \boldsymbol{\theta}_d) p(\mathbf{w}_{dn} | \mathbf{z}_{dn}, \boldsymbol{\beta}) ] \nonumber\\
& \times \prod_{d d'} p(\boldsymbol{\tau} | \boldsymbol{\theta}_d, \boldsymbol{\theta}_{d'}) p (s_{dd'} | \boldsymbol{\tau}_{dd'}, \boldsymbol{\mu}, \boldsymbol{\sigma})
\end{align}

\subsubsection{Regularization}
The basic idea of the regularization is
to align visual similarities with topic similarities
between two images.
To be specific,
it is to
classify the two images into
two categories
that show whether the two images
have the same semantic content,
and to align the classification result
from the topic distribution
with that from the visual content.

The latent variable $\boldsymbol{\tau} = [\tau_1~\tau_2]^T$ ,
called relational indicator,
is a $2$-dimensional binary-valued vector,
$\tau_1 + \tau_2 = 1$.
$\tau_1 = 1$ indicates that the two images
are regarded to have the same topics,
and otherwise,
the two images do not have the same topics.
It satisfies a multinomial distribution,
$\operatorname{Multinomial}(\operatorname{R}(\boldsymbol{\theta}_d, \boldsymbol{\theta}_{d'}))$,
where $\operatorname{R}(\boldsymbol{\theta}_d, \boldsymbol{\theta}_{d'}) =
[r_1(\boldsymbol{\theta}_d, \boldsymbol{\theta}_{d'}), r_2(\boldsymbol{\theta}_d, \boldsymbol{\theta}_{d'})]^T
=[r_{dd'1}~1-r_{dd'1}]^T$.
$r_{dd'1}$,
the probability of $\tau_1 = 1$,
is defined to describe the similarity
between two topics.
In essence,
$[r_1(\boldsymbol{\theta}_d, \boldsymbol{\theta}_{d'}), r_2(\boldsymbol{\theta}_d, \boldsymbol{\theta}_{d'})]^T$
reflects the probabilities
that the two images are recognized
to have the save topics or not.
For example,
$\operatorname{R}(\boldsymbol{\theta}_d, \boldsymbol{\theta}_{d'})$
can be defined as $[\operatorname{HI}(\boldsymbol{\theta}_d, \boldsymbol{\theta}_{d'}), 1 - \operatorname{HI}(\boldsymbol{\theta}_d, \boldsymbol{\theta}_{d'})]^T$,
where $\operatorname{HI}(\boldsymbol{\theta}_d, \boldsymbol{\theta}_{d'}) =
\sum_{i = 1}^K \min (\theta_{dk},\theta_{d'k})$ is a histogram intersection
over two topic distributions, $\boldsymbol{\theta}_d$ and $\boldsymbol{\theta}_{d'}$.

In this model,
$s_{dd'}$, an observed variable,
is the visual similarity
between two images $I_{d}$ and $I_{d'}$.
$\boldsymbol{\mu} = [\mu_1~\mu_2]^T$
and $\boldsymbol{\sigma} = [\sigma_1~\sigma_2]^T$ are 2-dimensional vectors,
and are used to describe two Gaussian distributions,
$\mathcal{N}(\mu_1, \sigma_1)$ and $\mathcal{N}(\mu_2, \sigma_2)$,
which correspond to the conditional probabilities
of the visual similarity,
conditioned on whether the two images have same semantic content.
It can be easily derived
that $\mu_1 > \mu_2$
since the larger the visual similarity,
the larger the probability that the two images have the same semantic content.

The probability of $s_{dd'}$,
conditioned on the topic distribution,
is given as follows,
\begin{align}
& p(s_{dd'} | \boldsymbol{\theta}_d, \boldsymbol{\theta}_{d'}, \boldsymbol{\mu}, \boldsymbol{\sigma}) \nonumber \\
= & \sum_{\boldsymbol{\tau}_{dd'}} p(s_{dd'} |\boldsymbol{\tau}_{dd'}, \boldsymbol{\mu}, \boldsymbol{\sigma})
p(\boldsymbol{\tau}_{dd'} | \boldsymbol{\theta}_d, \boldsymbol{\theta}_{d'}) \nonumber \\
= & r_1(\boldsymbol{\theta}_d, \boldsymbol{\theta}_{d'}) p(s_{dd'} | \mu_1, \sigma_1)
+ (1 - r_1(\boldsymbol{\theta}_d, \boldsymbol{\theta}_{d'})) p(s_{dd'} | \mu_2, \sigma_2).
\end{align}
We analyze the relation between visual similarity $s_{dd'}$ and
topic similarity $r_1(\boldsymbol{\theta}_d, \boldsymbol{\theta}_{d'})$
in a bidirectional way.
Given the topic distribution,
$\boldsymbol{\theta}_d$ and $\boldsymbol{\theta}_{d'}$,
we can obtain
\begin{align}
\operatorname{E}[s_{dd'} | \boldsymbol{\theta}_d, \boldsymbol{\theta}_{d'}, \boldsymbol{\mu}, \boldsymbol{\sigma}]
= & r_{dd'1} \mu_1 + (1 - r_{dd'1}) \mu_2.
\end{align}
This indicates that
the expectation of the visual similarity is larger
when the topic similarity is larger.
This is more reasonable
and more robust to noise,
compared with the direct requirement
that the visual similarity is larger
when the topic similarity is larger,
because of the gap
between visual contents and semantics.

Given the visual similarity, $s_{dd'}$,
the posterior that the two images have the same content
is computed by
\begin{align}
& P(\tau = 1|s_{dd'}) \nonumber \\
= & \frac{r_1(\boldsymbol{\theta}_d, \boldsymbol{\theta}_{d'}) p(s_{dd'} | \mu_1, \sigma_1)}
{r_1(\boldsymbol{\theta}_d, \boldsymbol{\theta}_{d'}) p(s_{dd'} | \mu_1, \sigma_1) +
(1 - r_1(\boldsymbol{\theta}_d, \boldsymbol{\theta}_{d'})) p(s_{dd'} | \mu_2, \sigma_2)}. \nonumber
\end{align}
Suppose we expect that
the two image have the same content
when $s_{dd'} \leqslant \bar{s}$.
This leads to that
\begin{align}
r_{dd'1} p(s_{dd'} | \mu_1, \sigma_1) >
(1 - r_{dd'1}) p(s_{dd'} | \mu_2, \sigma_2),
\end{align}
in the case $s_{dd'} \leqslant \bar{s}$.
This further means that
\begin{align}
r_{dd'1} > & \frac{p(s_{dd'} | \mu_2, \sigma_2)}
{p(s_{dd'} | \mu_1, \sigma_1) + p(s_{dd'} | \mu_2, \sigma_2)} \nonumber\\
> & \frac{1} {\frac{p(s_{dd'} | \mu_1, \sigma_1)}{p(s_{dd'} | \mu_2, \sigma_2)} + 1} \nonumber\\
> & \frac{1} {\min_{s_{dd'} \leqslant \bar{s}}\frac{p(s_{dd'} | \mu_1, \sigma_1)}{p(s_{dd'} | \mu_2, \sigma_2)} + 1}.
\end{align}
From the above analysis,
it can be concluded that
the topic similarity must be larger than some constant value
in order to
align it with the classification result from visual similarity.
This is a relaxant requirement
since it does not require that
the topic similarity must be larger
if the visual similarity is larger,
and hence also more reasonable
because there is some gap
between visual and semantic contents.

\begin{figure*}
\centering
\subfigure[(a)]{\label{fig:IllustrationRLDAInference:z}\includegraphics[scale = 1.5]{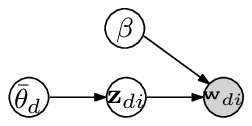}}~~~
\subfigure[(b)]{\label{fig:IllustrationRLDAInference:tau}\includegraphics[scale = 1.5]{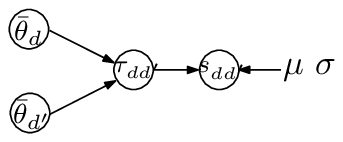}}~~~
\subfigure[(c)]{\label{fig:IllustrationRLDAInference:theta}\includegraphics[scale = 1.5]{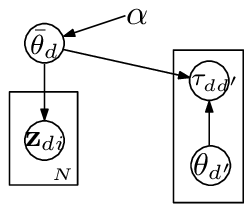}}
\caption{Illustration of the decomposition
for the inference scheme.
(a) shows the variables on which $\mathbf{z}$ depends,
(b) shows the variables on which $\boldsymbol{\tau}$ depends,
and (c) shows the variables on which $\boldsymbol{\theta}$ depends.}
\label{fig:IllustrationRLDAInference}
\end{figure*}

\subsubsection{Inference}
\label{sec:inference}
Let's first look at a possible solution,
the variational inference technique
that is used in LDA~\cite{BleiNJ03} to
estimate the posterior distribution of the latent variables:
\begin{align}
& p(\{\boldsymbol{\theta}_d\}, \{\mathbf{z}_{dn}\}_{dn},
\{\boldsymbol{\tau}_{dd'}\}_{dd'} |
\boldsymbol{\alpha}, \boldsymbol{\beta}, \boldsymbol{\mu}, \boldsymbol{\sigma};
\{\mathbf{w}_{dn}\}, \{\mathbf{s}_{dd'}\}) \nonumber \\
= & \frac{p(\{\boldsymbol{\theta}_d\}, \{\{\mathbf{w}_{dn}, \mathbf{z}_{dn}\}_n\}_d,
\{\mathbf{s}_{dd'}, \boldsymbol{\tau}_{dd'}\}_{dd'} |
\boldsymbol{\alpha}, \boldsymbol{\beta}, \boldsymbol{\mu}, \boldsymbol{\sigma})}
{p(\{\mathbf{w}_{dn}\}, \{\mathbf{s}_{dd'}\} | \boldsymbol{\alpha}, \boldsymbol{\beta}, \boldsymbol{\mu}, \boldsymbol{\sigma})}.
\end{align}

We introduce the following variational distribution
\begin{align}
& q(\{\boldsymbol{\theta}_d\}, \{\mathbf{z}_{dn}\}_{dn},
\{\boldsymbol{\tau}_{dd'}\}_{dd'} | \{\boldsymbol{\gamma}_d\}, \{\boldsymbol{\phi}_{dn}\}, \{\boldsymbol{\rho}_{dd'}\}) \nonumber \\
= & \prod_{d=1}^{M} [ q(\boldsymbol{\theta}_d | \boldsymbol{\gamma}_d)
\prod_{n=1}^{N_d} q(\mathbf{z}_{dn} | \boldsymbol{\phi}_{dn}) ]
\prod_{dd'} q(\boldsymbol{\tau}_{dd'} | \boldsymbol{\rho}_{dd'}),
\end{align}
where the Dirichlet parameter $\{\boldsymbol{\gamma}_d\}$,
the multinomial parameters $\{\boldsymbol{\phi}_{dn}\}$,
and the Dirichlet parameters $\{\boldsymbol{\rho}_{dd'}\}$
are the free variational parameters.
These variational parameters can be obtained
by solving the following optimization problem
\begin{align}
& \{\boldsymbol{\gamma}*_d\}, \{\boldsymbol{\phi}*_{dn}\}
\{\boldsymbol{\rho}*_{dd'}\}  \arg \min \operatorname{KL}(q || p).
\end{align}
This optimization problem can be solved
via an iterative fixed-point method.
For $\phi$ and $\rho$,
we can have the following update equations,
\begin{align}
\phi_{dni} & \propto \beta_{i\mathbf{w}_{dn}} \exp(\Psi(\gamma_{di}) - \Psi(\sum_{j=1}^K \gamma_{dj})), \\
\rho_{ddi} & \propto \exp(\operatorname{E}_q[\log {\operatorname{R}_i(\boldsymbol{\theta}_d, \boldsymbol{\theta}_{d'})}] ) + \frac{1}{\sigma_i\sqrt{2\pi}}\exp(-\frac{(s_{dd} - \mu_i)^2}{2\sigma_i^2})).
\end{align}
For $\boldsymbol{\gamma}$,
there is no closed-form solution.
The gradient decent based solution
requires the computation
of $\frac{\partial \operatorname{E}_q[\log \operatorname{R}_1(\boldsymbol{\theta}_d, \boldsymbol{\theta}_{d'})]}{\partial \gamma_{di}}$,
which is intractable.

In order to make the inference feasible,
instead,
we propose a hybrid sampling based approach,
which iteratively
samples two latent variables
$\mathbf{z}_{dn}$ and $\operatorname{\tau}_{dd'}$
and computes the conditional expectation,
$\bar{\boldsymbol{\theta}}_d  = \operatorname{E}(\boldsymbol{\theta}_d | \Theta - \{\boldsymbol{\theta}_d\})$:
\begin{enumerate}
  \item Sample $\mathbf{z}_{dn}$ from the conditional distribution
  $p(\mathbf{z}_{dn} | \Theta - \{\mathbf{z}_{dn}\}) \propto p(\mathbf{z}_{dn} | \bar{\boldsymbol{\theta}}_d) p(\mathbf{w}_{dn} | \mathbf{z}_{dn})$.
  This is depicted in Figure~\ref{fig:IllustrationRLDAInference:z}.
  \item Sample $\boldsymbol{\tau}_{dd'}$ from the conditional distribution
  $p(\boldsymbol{\tau}_{dd'} | \Theta - \{\boldsymbol{\tau}_{dd'}\}) \propto p(\boldsymbol{\tau}_{dd'} | \bar{\boldsymbol{\theta}}_d, \bar{\boldsymbol{\theta}}_{d'})
  p(s_{dd'} | \boldsymbol{\tau}_{dd'})$.
  This is depicted in Figure~\ref{fig:IllustrationRLDAInference:tau}.
  \item Compute the conditional expectation $\bar{\boldsymbol{\theta}}_d =
  \operatorname{E}(\boldsymbol{\theta}_d | \Theta - \{\boldsymbol{\theta}_d\})$.
  This is depicted in Figure~\ref{fig:IllustrationRLDAInference:theta}.
\end{enumerate}

From the definition,
$\mathbf{z}_{dn}$ is a discrete vector
with only one entry being $1$
and all the others being $0$,
thus sampling $\mathbf{z}_{dn}$
is straightforward.
Similarly,
sampling $\boldsymbol{\tau}_{dd'}$ is also straightforward.

We propose to
adopt the importance sampling approach
to
compute the conditional expectation $\bar{\boldsymbol{\theta}}_d$.
The conditional probability can be calculated
as below,
\begin{align}
p(\boldsymbol{\theta}_d | \Theta - \{\boldsymbol{\theta}_d\})
\propto p(\boldsymbol{\theta} | \boldsymbol{\alpha})
p(\{\mathbf{z}_{dn}\}_n | \boldsymbol{\theta}_d)
\prod_{d'} p(\boldsymbol{\tau}_{dd'} | \boldsymbol{\theta}_d, \boldsymbol{\theta}_{d'}).
\end{align}
We use $p(\boldsymbol{\theta} | \boldsymbol{\alpha} )$,
which can be easily sampled,
as the proposal function.
The conditional expectation is computed as follows,
\begin{align}
& \operatorname{E}(\boldsymbol{\theta}_d | \Theta - \{\boldsymbol{\theta}_d\}) \nonumber \\
= & \int \boldsymbol{\theta}_d p(\boldsymbol{\theta}_d | \Theta - \{\boldsymbol{\theta}_d\})d \boldsymbol{\theta}_d \nonumber \\
= & \frac{\int  \boldsymbol{\theta}_d p(\boldsymbol{\theta}_d | \boldsymbol{\alpha})
p(\{\mathbf{z}_{dn}\}_n | \boldsymbol{\theta}_d)
\prod_{d'} p(\boldsymbol{\tau}_{dd'} | \boldsymbol{\theta}_d, \boldsymbol{\theta}_{d'}) d \boldsymbol{\theta}_d}
{\int  p(\boldsymbol{\theta}_d | \boldsymbol{\alpha})
p(\{\mathbf{z}_{dn}\}_n | \boldsymbol{\theta}_d)
\prod_{d'} p(\boldsymbol{\tau}_{dd'} | \boldsymbol{\theta}_d, \boldsymbol{\theta}_{d'}) d \boldsymbol{\theta}_d} \nonumber \\
\approx &
\frac{\sum_{i=1}^{N} \boldsymbol{\theta}_d^{(i)}
p(\{\mathbf{z}_{dn}\}_n | \boldsymbol{\theta}_d^{(i)})
\prod_{d'} p(\boldsymbol{\tau}_{dd'} | \boldsymbol{\theta}_d^{(i)}, \boldsymbol{\theta}_{d'}) d \boldsymbol{\theta}_d}
{\sum_{i=1}^{N}
p(\{\mathbf{z}_{dn}\}_n | \boldsymbol{\theta}_d^{(i)})
\prod_{d'} p(\boldsymbol{\tau}_{dd'} | \boldsymbol{\theta}_d^{(i)}, \boldsymbol{\theta}_{d'}) d \boldsymbol{\theta}_d}.
\end{align}

\subsubsection{Parameter estimation}
\label{sec:parameterestimation}
Given the expectations $\{\bar{\boldsymbol{\theta}}_d\}$
of all the documents,
$\boldsymbol{\alpha}$ can be estimated
by maximizing the likelihood,
\begin{align}
p(\{\bar{\boldsymbol{\theta}}_d\}; \boldsymbol{\alpha})
= \prod_d \frac{\Gamma (\sum_k \alpha_k)}{\prod_k \Gamma(\alpha_k)} \prod_k \bar{\theta}_k^{\alpha_k - 1}.
\end{align}
The maximization problem can be solved
by a fixed-point iteration~\cite{Minka09}.

Given samples $\{\mathbf{z}_{dn}\}_{dn}$,
$\boldsymbol{\beta}$ can also be estimated
by maximizing the likelihood,
\begin{align}
p(\{\mathbf{w}_{dn}\}_{dn}|\{\mathbf{z}_{dn}\}_{dn}; \boldsymbol{\beta})
= \prod_{dn} p(\mathbf{w}_{dn}|\mathbf{z}_{dn}; \boldsymbol{\beta}).
\end{align}
Here, $\boldsymbol{\beta}$ can be regarded as a Markov matrix
from $\mathbf{z}$ to $\mathbf{w}$
that can be easily computed.

Given samples $\boldsymbol{\tau}_{dd'}$,
the Gaussian parameters,
$\boldsymbol{\mu}_1$, $\boldsymbol{\sigma}_1$,
$\boldsymbol{\mu}_2$ and $\boldsymbol{\sigma}_2$,
can be easily estimated
from visual similarities $s_{dd'}$.

In a summary,
given the observations,
tags $\mathcal{W}$ associated with
visual similarities $\{s_{dd'}\}$,
the whole algorithm is an iterative scheme
in which each iteration consists of
latent variable inference (Section~\ref{sec:inference})
and model parameter estimation (this section).

\subsubsection{Tag relevance}
In the rLDA, we can estimate the tag relevance
jointly using the information from the image
and the information from other images.
The relevance of a tag $\mathbf{w}$ for one image
is formulated as the probability
conditioned on the set of tags $\mathcal{W}_d$
associated with this image,
and other sets of tags $\mathcal{W} - \mathcal{W}_d$.
It is mathematically formulated as
\begin{align}
& p_d(\mathbf{w} | \{\mathcal{W}\}_{d=1}^M, \{I_d\}_{d=1}^M) \nonumber \\
= &
\sum_{\bar{\boldsymbol{z}}} p(\mathbf{w}, \bar{\boldsymbol{z}}_d | \{\mathcal{W}\}_{d=1}^M, \{I_d\}_{d=1}^M) \nonumber \\
= &\sum_{\bar{\boldsymbol{z}}} p(\mathbf{w} | \bar{\boldsymbol{z}}) p(\bar{\boldsymbol{z}} | \{\mathcal{W}\}_{d=1}^M, \{I_d\}_{d=1}^M) \nonumber \\
= &\sum_{\bar{\boldsymbol{z}}} p(\mathbf{w} | \bar{\boldsymbol{z}}) \int p(\bar{\boldsymbol{z}}, \boldsymbol{\theta}_d | \{\mathcal{W}\}_{d=1}^M, \{I_d\}_{d=1}^M) d \boldsymbol{\theta}_d \nonumber \\
= &\sum_{\bar{\boldsymbol{z}}} p(\mathbf{w} | \bar{\boldsymbol{z}}) \int p(\bar{\boldsymbol{z}} | \boldsymbol{\theta}_d) p(\boldsymbol{\theta}_d | \{\mathcal{W}\}_{d=1}^M, \{I_d\}_{d=1}^M) d \boldsymbol{\theta}_d \nonumber \\
\approx &\sum_{\bar{\boldsymbol{z}}} p(\mathbf{w} | \bar{\boldsymbol{z}})  p(\bar{\boldsymbol{z}} | \bar{\boldsymbol{\theta}}_d).
\end{align}
The computation of tag relevance is similar to that in LDA.
Differently, this approximation is obtained
by jointly considering the information
from the other images.
The tag relevance model is illustrated using a graphical representation
in~Figure~\ref{fig:IllustrationByDeeplearning:rlda}.

\section{Experiments}
\subsection{Setup}
\subsubsection{Dataset}
Our experiments are conducted
over two datasets:
the MSRA-TAG dataset~\cite{LiuHYWZ09}
and the NUS-WIDE-LITE dataset~\cite{ChuaTHLLZ09}.
The MSRA-TAG dataset consists of $50,000$ images and their associated tags
that are downloaded from Flickr
using ten popular tags,
including cat, automobile, mountain, water, sea, bird, tree, sunset, flower and sky.
$13,330$ distinctive tags are obtained
after filtering out the misspelling and meaningless tags.
Similar to~\cite{LiuHYWZ09},
for quantitative evaluation,
$10,000$ images are randomly selected
from the dataset and manually labeled
to build the ground truth.
For each image, we ask volunteers to mark the relevance of each tag with a score,
ranging from $1$ (the least relevant) to $5$ (the most relevant).
We perform the tag reranking task over this dataset.

The NUS-WIDE-LITE dataset~\cite{ChuaTHLLZ09} consists of $27,818$ images
with tags provided by users.
$1,000$ filtered tags that appeared the most frequently
were used.
We perform the image rettagging task over this dataset.
In our experiments, the top $5$ tags with the highest scores
are chosen as the new tags of the image
and the performance of algorithms is evaluated
on $81$ tags
where the ground truth for these tags are provided in~\cite{ChuaTHLLZ09}.

\subsubsection{Evaluation}
The task of tag reranking is
to rank the original tags of an image according to their relevances.
The confidence scores for the tags are produced
and are used to rerank the tags.
The normalized discounted cumulative gain (NDCG) measurement is adopted
as the evaluation measure, which is calculated as
$\operatorname{NDCG}_n = Z_n\sum_{i=1}^n(2^{r(i)}-1)/\log(1+i)$,
where $r(i)$ is the relevance score of the $i$-th tag
and $Z_n$ is a normalization constant
that is chosen so that the NDCG score of the optimal ranking is 1.
We use the average of the NDCG scores of all the images
to compare the performance.

In the task of image retagging, each image is assigned a number of tags.
The assigned tags can be either original tags
or new tags that are not among the originals.
The top $5$ tags with the highest scores are chosen
as the retagging results of the image.
We have the ground truth of $81$ tags,
where whether each image in the dataset is related to these tags is provided.
Similar to~\cite{LiuYHZ11},
we perform a retrieval process based on retagging results
and evaluate the average F-measure of $81$ tags to
compare the performance.
The F-measure is calculated as $\frac{2 \times \operatorname{precision} \times \operatorname{recall}}{\operatorname{precision}+\operatorname{recall}}$,
where $\operatorname{precision}$ and $\operatorname{recall}$
are computed form the returned retrieval list.

\subsection{Methods}
To evaluate the performance of our approach, regularized LDA (rLDA),
for tag refinement,
we also report the experimental results of
existing state-of-the-art approaches.
\begin{enumerate}
  \item BaseLine. The score is computed
  based on the original tag ranking provided by users according to the uploading time.

  \item Random walk with restart (RWR)~\cite{WangJZZ06}.
  This method performs a random walk process on a graph
  that encodes the relationship between tags.
  It only uses the text information without
  using the visual information.

  \item Tag ranking based on visual and semantic consistency (TRVSC)~\cite{LiuHYWZ09}.
  This work follows RWR~\cite{WangJZZ06} using a random walk based method.
  The difference is that
  when constructing the graph over tags
  TRVSC considers the visual similarity between images.

  \item Tag refinement based on low-rank, content-tag prior and error sparsity (LRCTPES)~\cite{ZhuYM10}.
  This approach formulates the tag refinement problem
  as a decomposition of the user-provided tag matrix
  into a low-rank matrix and a sparse error matrix,
  targeting the optimality
  by low-rank, content consistency,
  tag correlation and error sparsity.

  \item Collaborative retagging (CRT)~\cite{LiuYHZ11}.
  The CRT process is formulated as a multiple graph-based multi-label learning problem.
  This work also proposes a tag-specific visual sub-vocabulary learning method.
  In our implementation we did not use the sub-vocabulary part
  because we focus mainly on the approach of rag refinement,
  not feature extraction.
  The parameters of CRT are tuned using the grid search method in~\cite{LiuYHZ11}.

  \item Separate retagging (SRT).
  SRT performs the same method
  as CRT~\cite{LiuYHZ11},
  but without considering tag similarity.
  We also use the grid search method described in~\cite{LiuYHZ11} to find the best parameters.

  \item Latent Dirichlet Allocation (LDA)~\cite{BleiNJ03}.
  We perform LDA
  by considering each image as a document
  and each tag as a word.

\end{enumerate}

In our experiments,
all the approaches use the same visual features,
a $225$-dimensional block-wise color moment.
The visual similarity is calculated as $s_{ij} = \exp\{-\frac{\|\mathbf{f}_i-\mathbf{f}_j\|_2^2}{2\gamma}\}$,
where $\mathbf{f}_i$ is the visual feature of image $I_i$
and $\gamma = 9\operatorname{E}[\|\mathbf{f}_i-\mathbf{f}_j\|_2^2]$
with $\operatorname{E}$ being the expectation operator.
Two images are connected in the rLDA model
only when their similarity is higher than $0.2$.

\subsection{Results on tag reranking}
\subsubsection{Comparison}

\begin{figure}[t]
\centering
\includegraphics[width=0.45\textwidth]{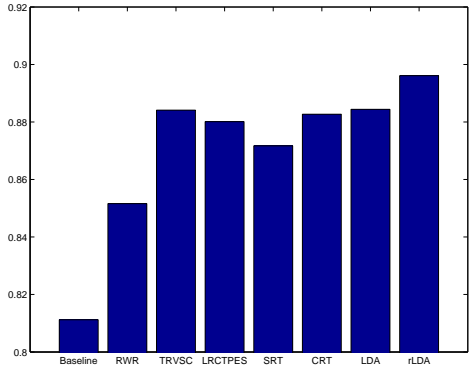}
\caption{Performance comparison for tag reranking.
The horizontal axis corresponds to different methods,
and the vertical axis corresponds to the NDCG score.}
\label{fig:tagranking}
\end{figure}

The results of tag reranking are
reported in Figure~\ref{fig:tagranking}.
It can be observed that
all approaches perform better results than the baseline result.
It demonstrates that tag refinement is
a useful process.
Among these methods,
RWR and LDA are based on the tag information and
do not take into account of the visual information.
SRT only uses the visual information.
TRVSC, LRCTPES, CRT and rLDA make use of both the textual and visual information.
We can see that both the textual information
and the visual information can make great contributions to tag refinement.
Though LDA does not use the visual information,
it outperforms most of other methods,
showing the significant benefit of
jointly estimating the tag similarity and the tag relevance
and the powerfulness of exploring multi-wise relationships
using the topic model.
Our approach, rLDA,
further improves LDA by encouraging visually similar images
having similar topic distributions.
The superiority of rLDA over LDA
justifies our analysis
that rLDA is deeper than LDA illustrated
in Figure~\ref{fig:IllustrationByDeeplearning}.

The superior performance of our approach
can be justified from the deep learning theory~\cite{HintonOT06},
which shows that a deeper network has large potentials
to achieve better performance.
By comparison,
the random walk based approaches essentially
use shallow structures,
which only consists of two levels,
the provided tags as the input level
and the tag being considered as the output level.
The LDA based approach is with a deep structure,
introducing a latent topic level,
which has potential to get better performance.
The proposed regularized LDA model is deeper,
with four levels,
the tags associated with other images
as the first level,
the latent topics of other images
and the tags of the image being considered
as the second level,
the latent topic as the third level,
and the tag being considered as the output level.
The comparison has been illustrated
in Figure~\ref{fig:IllustrationByDeeplearning}.

\subsubsection{Empirical analysis}
\begin{figure}[t]
\centering
\includegraphics[width=0.45\textwidth]{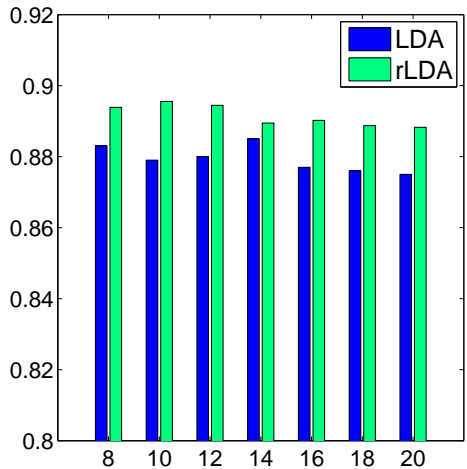}
\caption{Performance comparison
of rLDA and LDA
with different numbers of topics.}
\label{fig:rldavslda}
\end{figure}
\begin{figure}[t]
\centering
\includegraphics[width=0.45\textwidth]{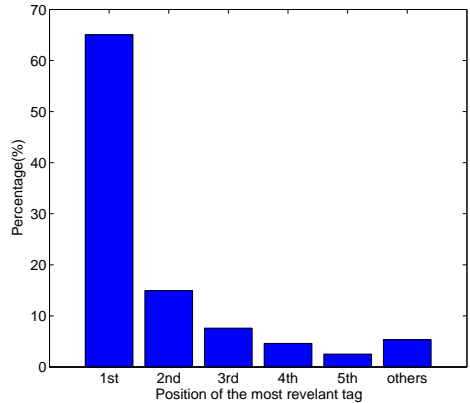}
\caption{The statistics
of the position at which the truly most relevant tag is ranked
for our approach.
The horizontal axis corresponds to the position
and the vertical axis corresponds to the percentage of the images.}
\label{fig:rldaposition}
\end{figure}


To illustrate the superiority of rLDA over LDA clearer,
we compare their performances
using different numbers of topics, $K$,
which are shown in Figure~\ref{fig:rldavslda}.
We have at least two observations.
The first one is that
taking visual information into account can be effective for the tag refinement task
from the fact that rLDA consistently outperforms LDA on different number of topics.
The second one is that the performances of both methods begins to decrease
when $K$ grows too large.
This is reasonable.
Considering the extreme case that $K \geqslant V$,
it can be validated
that it will overfit the data distribution
if setting each word as a single topic,
which indicates that the relations among tags tend to be useless
when $K$ is too large.

\begin{figure*}
{\footnotesize
\begin{tabular}{p{.1\textwidth}p{.2\textwidth}p{.2\textwidth}p{.2\textwidth}p{.2\textwidth}}
&\includegraphics[width = .2\textwidth, height = .2\textwidth, clip]{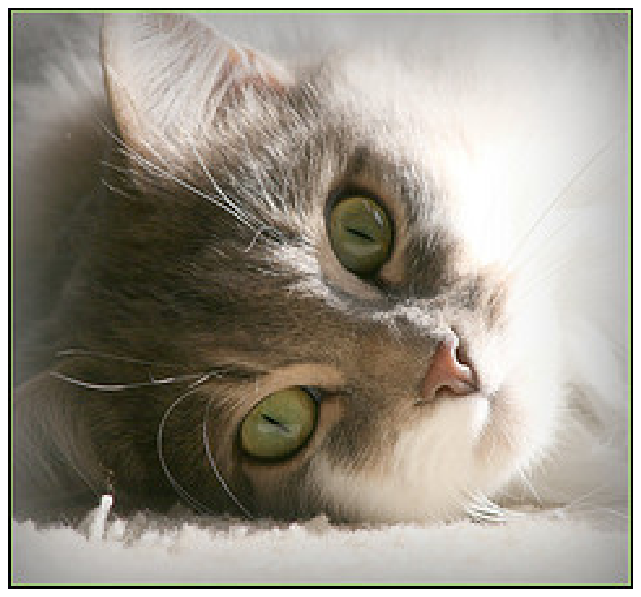}
& \includegraphics[width = .2\textwidth, height = .2\textwidth, clip]{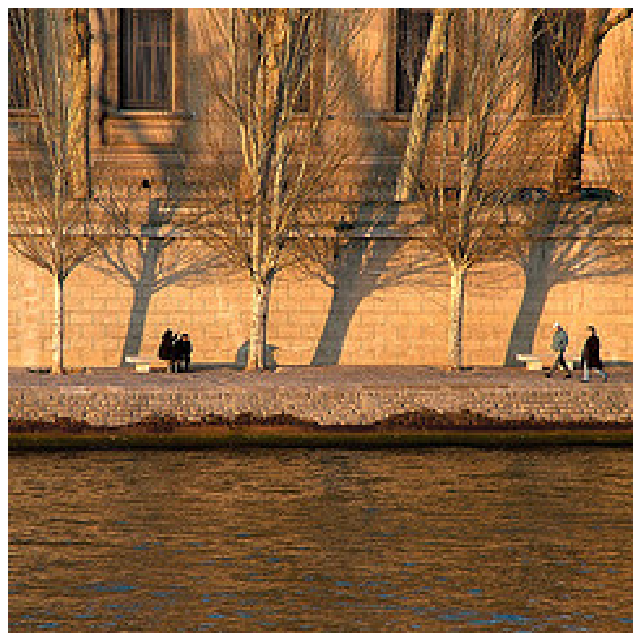}
& \includegraphics[width = .2\textwidth, height = .2\textwidth, clip]{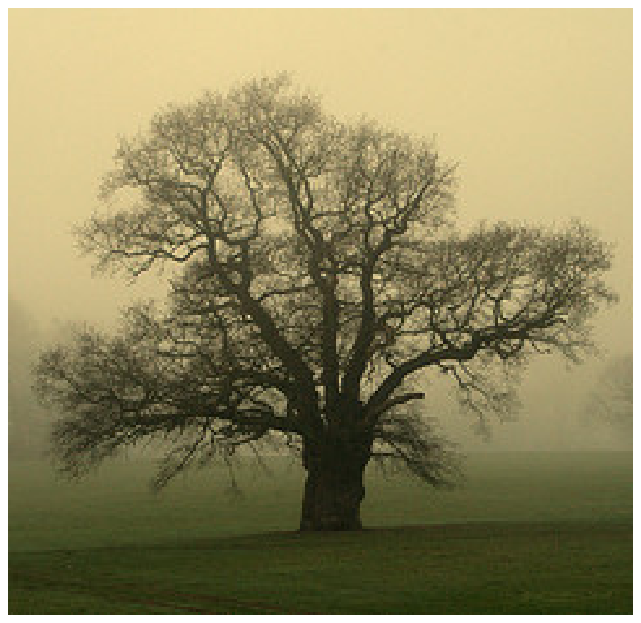}
& \includegraphics[width = .2\textwidth, height = .2\textwidth, clip]{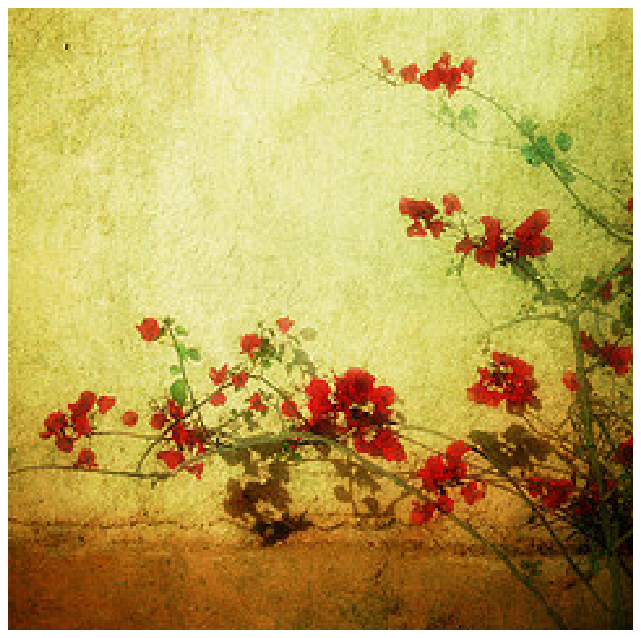} \\
Original tags      &
family winter friends portrait art nature cat nose australia
& friends light sunlight paris france water
& park winter england cold tree nature
& travel summer sun flower nature greece
\\
Refined tags&
cat nature portrait nose winter art friends family australia
& water light france paris sunlight friends
& tree nature winter england park cold
& flower nature sun travel summer greece
\\

&\includegraphics[width = .2\textwidth, height = .2\textwidth, clip]{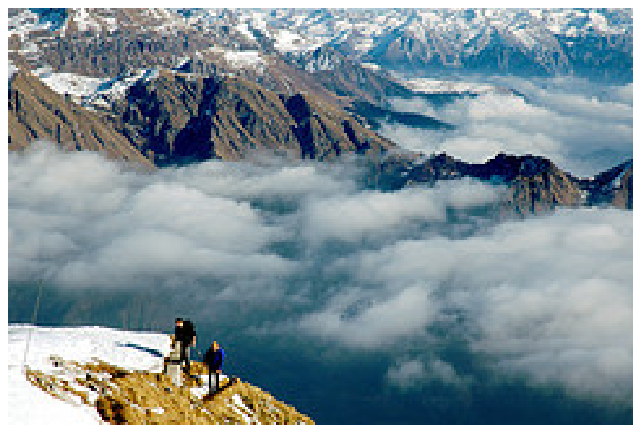}
& \includegraphics[width = .2\textwidth, height = .2\textwidth, clip]{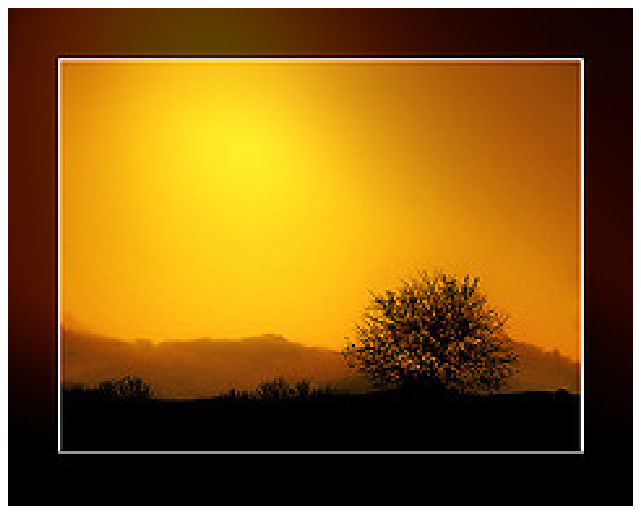}
& \includegraphics[width = .2\textwidth, height = .2\textwidth, clip]{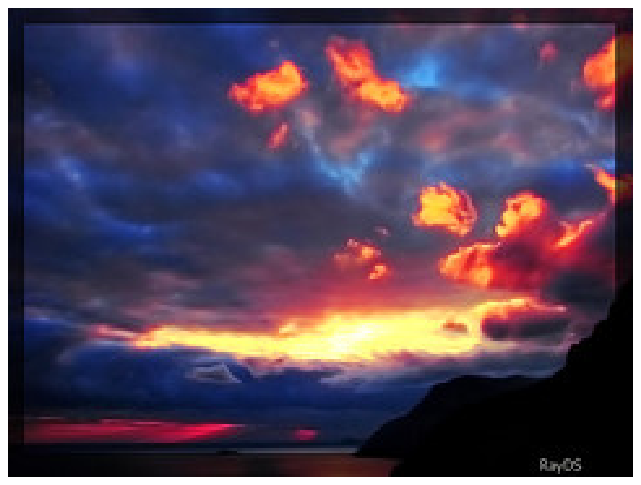}
& \includegraphics[width = .2\textwidth, height = .2\textwidth, clip]{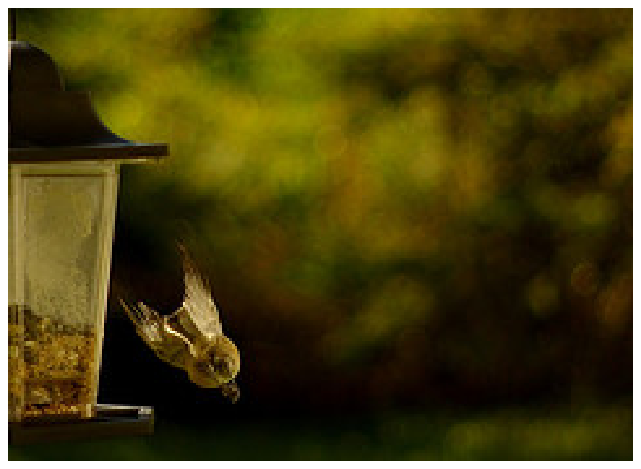}\\
Original tags&
winter sea italy mountain snow alps weather landscape
& travel sunset canada tree nature quebec quality
& travel sea sky italy landscape sony cielo
& england italy cold holland bird grey flight
\\
Refined tags&
landscape mountain sea snow winter italy alps weather
& sunset tree nature travel quality canada quebec
& sky landscape sea travel italy cielo sony
& bird flight italy england holland cold grey
\end{tabular}
}
\caption{Tag ranking examples.}
\label{fig:tagrakingexamples}
\end{figure*}

To understand our approach more deeply,
we report the percentage of the images in which the truly most relevant tag is ranked in different positions.
in Figure~\ref{fig:rldaposition}.
We can see that over 60 percent of the images have their most relevant tag at the first position.
This can be helpful for related works like image retrieval, group recommendation, etc.
Some examples of refined tags are depicted in Figure~\ref{fig:tagrakingexamples}.

\subsection{Results on image retagging}
\begin{figure}[t]
\centering
\includegraphics[width=0.45\textwidth]{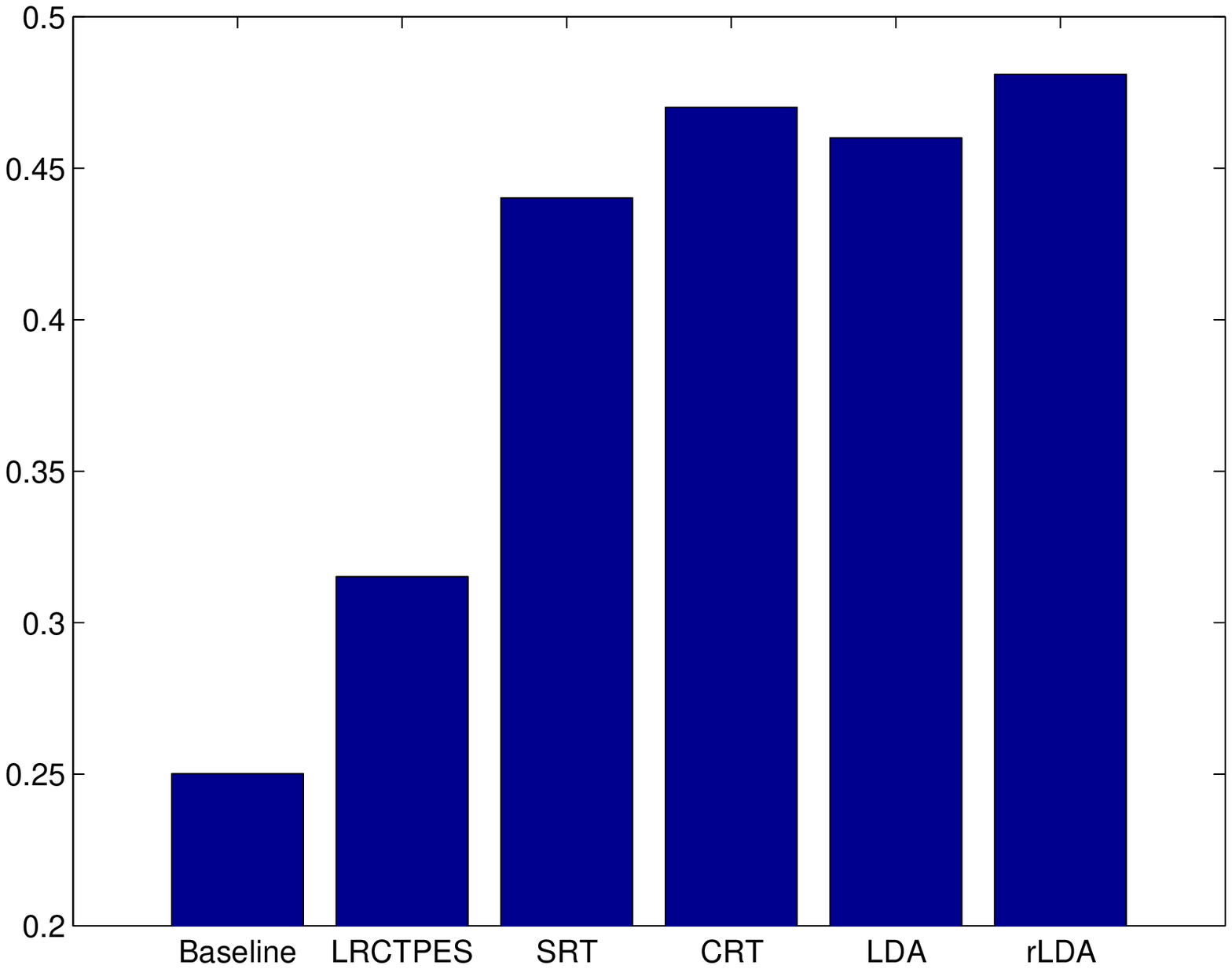}
\caption{The results of retagging on the NUS-WIDE-LITE dataset.
The vertical axis corresponds to F-measure.}
\label{fig:retag}
\end{figure}

Different from tag reranking,
retagging~\cite{LiuYHZ11} aims to suggest a set of tags
that are assigned according to the original tags.
These tags may not necessarily be contained
in the original tags.
In this task, the results of five methods,
SRT, LRCTPES, CRT, LDA and our approach, are reported.
Other methods based on random walks only produce scores for original tags
and can not perform the retagging task
Figure~\ref{fig:retag}~shows the experiment results.

The retagging results of all methods outperformed the base line
that is based on the original tag list.
This demonstrates that image retagging can make significant contributions for image retrieval.
In the image retagging task LDA does not perform as good as
in the tag reranking task.
This is because
LDA is unable to deal with the images without tags provided by users,
while methods using visual features
can tag the images using the tags of similar images.
Improving LDA with using visual content of images,
our method, rLDA, gets the best result.
In both tag reranking and image retagging tasks,
the proposed method performs the best, which is because our model is based on a deeper structure
and can exploit the semantic information derived from the topic level.

\section{Conclusion}
This paper presents a regularized latent Dirichlet allocation approach
for tag refinement.
Our approach succeeds from
the factors:
(1) Our approach explores the multi-wise relationship
among the tags
that are mined from both textual and visual information;
(2) Our approach explores a deep structure
that has large capability to refine tags.
Experimental results also demonstrate the superiority
of our approach over existing state-of-the-art approaches
for tag refinement.

\small{
\bibliographystyle{abbrv}
\bibliography{tag}

\begin{thebibliography}{10}

\bibitem{AmesN07}
M.~Ames and M.~Naaman.
\newblock Why we tag: motivations for annotation in mobile and online media.
\newblock In {\em CHI}, pages 971--980, 2007.

\bibitem{BarnardDFFBJ03}
K.~Barnard, P.~Duygulu, D.~A. Forsyth, N.~de~Freitas, D.~M. Blei, and M.~I.
  Jordan.
\newblock Matching words and pictures.
\newblock {\em Journal of Machine Learning Research}, 3:1107--1135, 2003.

\bibitem{Bengio09}
Y.~Bengio.
\newblock Learning deep architectures for ai.
\newblock {\em Foundations and Trends in Machine Learning}, 2(1):1--127, 2009.

\bibitem{BleiJ03}
D.~M. Blei and M.~I. Jordan.
\newblock Modeling annotated data.
\newblock In {\em SIGIR}, pages 127--134, 2003.

\bibitem{BleiNJ03}
D.~M. Blei, A.~Y. Ng, and M.~I. Jordan.
\newblock Latent dirichlet allocation.
\newblock {\em Journal of Machine Learning Research}, 3:993--1022, 2003.

\bibitem{BundschusYTRD09}
M.~Bundschus, S.~Yu, V.~Tresp, A.~Rettinger, and M.~Dejori.
\newblock Hierarchical bayesian models for collaborative tagging systems.
\newblock In {\em ICDM}, 2009.

\bibitem{CarneiroCMV07}
G.~Carneiro, A.~B. Chan, P.~J. Moreno, and N.~Vasconcelos.
\newblock Supervised learning of semantic classes for image annotation and
  retrieval.
\newblock {\em IEEE Trans. Pattern Anal. Mach. Intell.}, 29(3):394--410, 2007.

\bibitem{ChangB09}
J.~Chang and D.~M. Blei.
\newblock Relational topic models for document networks.
\newblock {\em Journal of Machine Learning Research - Proceedings Track},
  5:81--88, 2009.

\bibitem{ChenCCTHW08}
H.-M. Chen, M.-H. Chang, P.-C. Chang, M.-C. Tien, W.~H. Hsu, and J.-L. Wu.
\newblock Sheepdog: group and tag recommendation for flickr photos by automatic
  search-based learning.
\newblock In {\em ACM Multimedia}, pages 737--740, 2008.

\bibitem{ChuaTHLLZ09}
T.-S. Chua, J.~Tang, R.~Hong, H.~Li, Z.~Luo, and Y.~Zheng.
\newblock Nus-wide: a real-world web image database from national university of
  singapore.
\newblock In {\em CIVR}, 2009.

\bibitem{DattaJLW08}
R.~Datta, D.~Joshi, J.~Li, and J.~Z. Wang.
\newblock Image retrieval: Ideas, influences, and trends of the new age.
\newblock {\em ACM Comput. Surv.}, 40(2), 2008.

\bibitem{FengL08}
Y.~Feng and M.~Lapata.
\newblock Automatic image annotation using auxiliary text information.
\newblock In {\em ACL}, pages 272--280, 2008.

\bibitem{HintonOT06}
G.~E. Hinton, S.~Osindero, and Y.~W. Teh.
\newblock A fast learning algorithm for deep belief nets.
\newblock {\em Neural Computation}, 18(7):1527--1554, 2006.

\bibitem{JeonM04}
J.~Jeon and R.~Manmatha.
\newblock Using maximum entropy for automatic image annotation.
\newblock In {\em CIVR}, pages 24--32, 2004.

\bibitem{JiangCL06}
W.~Jiang, S.-F. Chang, and A.~C. Loui.
\newblock Active context-based concept fusionwith partial user labels.
\newblock In {\em ICIP}, pages 2917--2920, 2006.

\bibitem{JinCS04}
R.~Jin, J.~Y. Chai, and L.~Si.
\newblock Effective automatic image annotation via a coherent language model
  and active learning.
\newblock In {\em ACM Multimedia}, pages 892--899, 2004.

\bibitem{KennedyCK06}
L.~S. Kennedy, S.-F. Chang, and I.~Kozintsev.
\newblock To search or to label?: predicting the performance of search-based
  automatic image classifiers.
\newblock In {\em Multimedia Information Retrieval}, pages 249--258, 2006.

\bibitem{KrestelF09}
R.~Krestel and P.~Fankhauser.
\newblock Tag recommendation using probabilistic topic models.
\newblock In {\em ECML/PKDD Discovery Challenge (DC'09), Workshop at ECML/PKDD
  2009)}, 2009.

\bibitem{KrestelFN09}
R.~Krestel, P.~Fankhauser, and W.~Nejdl.
\newblock Latent dirichlet allocation for tag recommendation.
\newblock In {\em RecSys}, pages 61--68, 2009.

\bibitem{LiW03}
J.~Li and J.~Z. Wang.
\newblock Automatic linguistic indexing of pictures by a statistical modeling
  approach.
\newblock {\em IEEE Trans. Pattern Anal. Mach. Intell.}, 25(9):1075--1088,
  2003.

\bibitem{LiSW08}
X.~Li, C.~G.~M. Snoek, and M.~Worring.
\newblock Learning tag relevance by neighbor voting for social image retrieval.
\newblock In {\em Multimedia Information Retrieval}, pages 180--187, 2008.

\bibitem{LiSW09}
X.~Li, C.~G.~M. Snoek, and M.~Worring.
\newblock Learning social tag relevance by neighbor voting.
\newblock {\em IEEE Transactions on Multimedia}, 11(7):1310--1322, 2009.

\bibitem{LiuHWZ10}
D.~Liu, X.-S. Hua, M.~Wang, and H.-J. Zhang.
\newblock Image retagging.
\newblock In {\em ACM Multimedia}, pages 491--500, 2010.

\bibitem{LiuHYWZ09}
D.~Liu, X.-S. Hua, L.~Yang, M.~Wang, and H.-J. Zhang.
\newblock Tag ranking.
\newblock In {\em WWW}, pages 351--360, 2009.

\bibitem{LiuYHZ11}
D.~Liu, S.~Yan, X.-S. Hua, and H.-J. Zhang.
\newblock Image retagging using collaborative tag propagation.
\newblock {\em IEEE Transactions on Multimedia}, 13(4):702--712, 2011.

\bibitem{Miller95}
G.~A. Miller.
\newblock Wordnet: A lexical database for english.
\newblock {\em Commun. ACM}, 38(11):39--41, 1995.

\bibitem{Minka09}
T.~P. Minka.
\newblock Estimating a dirichlet distribution.
\newblock Technical report, 2009.

\bibitem{NguyenKPT10}
C.-T. Nguyen, N.~Kaothanthong, X.~H. Phan, and T.~Tokuyama.
\newblock A feature-word-topic model for image annotation.
\newblock In {\em CIKM}, pages 1481--1484, 2010.

\bibitem{PutthividhyaAN10}
D.~Putthividhya, H.~T. Attias, and S.~S. Nagarajan.
\newblock Supervised topic model for automatic image annotation.
\newblock In {\em ICASSP}, pages 1894--1897, 2010.

\bibitem{QiHRTMZ07}
G.-J. Qi, X.-S. Hua, Y.~Rui, J.~Tang, T.~Mei, and H.-J. Zhang.
\newblock Correlative multi-label video annotation.
\newblock In {\em ACM Multimedia}, pages 17--26, 2007.

\bibitem{SigurbjornssonZ08}
B.~Sigurbj{\"o}rnsson and R.~van Zwol.
\newblock Flickr tag recommendation based on collective knowledge.
\newblock In {\em WWW}, pages 327--336, 2008.

\bibitem{WangJZZ06}
C.~Wang, F.~Jing, L.~Zhang, and H.~Zhang.
\newblock Image annotation refinement using random walk with restarts.
\newblock In {\em ACM Multimedia}, pages 647--650, 2006.

\bibitem{WangHLZYC12}
M.~Wang, R.~Hong, G.~Li, Z.-J. Zha, S.~Yan, and T.-S. Chua.
\newblock Event driven web video summarization by tag localization and key-shot
  identification.
\newblock {\em IEEE Transactions on Multimedia}, 14(4):975--985, 2012.

\bibitem{WangNHC12}
M.~Wang, B.~Ni, X.-S. Hua, and T.-S. Chua.
\newblock Assistive tagging: A survey of multimedia tagging with human-computer
  joint exploration.
\newblock {\em ACM Comput. Surv.}, 44(4):25, 2012.

\bibitem{WeinbergerSZ08}
K.~Q. Weinberger, M.~Slaney, and R.~van Zwol.
\newblock Resolving tag ambiguity.
\newblock In {\em ACM Multimedia}, pages 111--120, 2008.

\bibitem{WuYYH09}
L.~Wu, L.~Yang, N.~Yu, and X.-S. Hua.
\newblock Learning to tag.
\newblock In {\em WWW}, pages 361--370, 2009.

\bibitem{XuWHL09}
H.~Xu, J.~Wang, X.-S. Hua, and S.~Li.
\newblock Tag refinement by regularized lda.
\newblock In {\em ACM Multimedia}, pages 573--576, 2009.

\bibitem{ZhouCQX11}
N.~Zhou, W.~K. Cheung, G.~Qiu, and X.~Xue.
\newblock A hybrid probabilistic model for unified collaborative and
  content-based image tagging.
\newblock {\em IEEE Trans. Pattern Anal. Mach. Intell.}, 33(7):1281--1294,
  2011.

\bibitem{ZhuYM10}
G.~Zhu, S.~Yan, and Y.~Ma.
\newblock Image tag refinement towards low-rank, content-tag prior and error
  sparsity.
\newblock In {\em ACM Multimedia}, pages 461--470, 2010.

\end{thebibliography}
}

\end{document}